\documentclass[aps,prl,twocolumn,superscriptaddress]{revtex4}
\usepackage{natbib}
\usepackage{amsmath}
\usepackage{amssymb}
\usepackage{amsthm}
\usepackage{graphicx}
\usepackage{graphics}
\usepackage{subfigure}
\usepackage{bm}
\usepackage{hyperref}
\usepackage{epsfig}

\def\up{\left|\uparrow\right\rangle}
\def\down{\left|\downarrow\right\rangle}

\def\v{v_{\scriptscriptstyle\Delta}}

\begin{document}

\title{Predicted signatures of $p$-wave superfluid phases and Majorana zero modes of fermionic atoms in RF absorption}%

\author{Eytan Grosfeld}
\affiliation{Department of Condensed Matter Physics, Weizmann
Institute of Science, Rehovot 76100, Israel}

\author{Nigel R. Cooper}
\affiliation{T.C.M. Group, Cavendish Laboratory, J.J. Thomson Avenue, Cambridge, CB3
0HE, United Kingdom}

\author{Ady Stern}
\affiliation{Department of Condensed Matter Physics, Weizmann
Institute of Science, Rehovot 76100, Israel}

\author{Roni Ilan}
\affiliation{Department of Condensed Matter Physics, Weizmann
Institute of Science, Rehovot 76100, Israel}

\date{\today}

\begin{abstract}

We study the superfluid phases of quasi-2D atomic Fermi gases interacting via
a $p$-wave Feshbach resonance. We calculate the absorption spectra
of these phases under a hyperfine transition, for both non-rotating and rotating superfluids.
We show that one can identify the different
phases of the $p$-wave superfluid from the absorption
spectrum. The absorption spectrum shows clear signatures of the
existence of Majorana zero modes at the cores of vortices of the
weakly-pairing $p_x+ip_y$ phase.

\end{abstract}

\pacs{}

\maketitle

% ----------------------------------------------------------------

The observation of $p$-wave Feshbach resonances in cold fermion
gases\cite{expt} opens up the possibility of the experimental
realisation of unconventional superfluid states with non-zero pairing
angular momentum\cite{theory,gurarie,cheng}.
It has recently been predicted\cite{gurarie,cheng} that a
one-component fermion gas interacting via a $p$-wave Feshbach
resonance shows a series of $p$-wave superfluid phases, including
a $p_x$ phase, and both strong- and weak- pairing
phases\cite{readgreen} of $p_x\pm i p_y$ symmetry.
The strong- and weak- pairing phases differ in the nature of their
excitation spectra, and are separated by a topological phase
transition\cite{volovik,readgreen}.

Among these, the weak-pairing $p_x\pm ip_y$ phase has recently
received considerable theoretical attention.  The vortex
excitations of this phase are predicted to hold gapless Majorana
fermions on their cores, which should give rise to {\it
non-abelian} exchange statistics in a quasi 2D
geometry\cite{mooreread,readgreen}. Such excitations may have
their uses in topologically-protected quantum computation
\cite{kitaev1,gurarie,tewari}.  However, to our knowledge, there as yet
are no experimental observations of these very interesting
physical properties in any realisation of a $p_x\pm ip_y$
superfluid.  Developing methods to experimentally probe the
properties of $p$-wave superfluid phases in cold atom systems is
then of high and timely interest.

In this Letter we show that features in the RF absorption spectrum of
a fermionic gas provide unambiguous signatures of the different
$p$-wave superfluid phases.
We show that the weak-pairing phases can be identified from the
strong-pairing phases in the absorption spectrum of the superfluid at
rest.
Most significantly, we focus on the $p_x+i p_y$ phase and show that for a {\it rotating} fluid at
weak-pairing, the absorption spectrum has clear and
striking signatures of the zero energy Majorana fermion modes on the
cores of the vortices.  This spectrum is therefore a direct probe of
the physics underlying the proposed non-abelian exchange statistics of
this phase.  (These experimental signatures do not require
manipulation of positions of vortices or the use of other localised
probes\cite{tewari}.)
Furthermore, we show that the absorption spectrum for the
rotated $p_x+ ip_y$ weak-pairing phase depends on the {\it sense} of
the rotation.  This dependence is a direct indication of the
time-reversal symmetry breaking in this phase.

We consider a gas of identical fermionic atoms in an internal state
that we denote $\down$, interacting via a $p$-wave Feshbach
resonance. The phase diagram of the system has the same qualitative
form in wide~\cite{cheng} and narrow~\cite{gurarie} resonance limits.
Depending on temperature, and on the detuning and anisotropic
splitting of the Feshbach resonance, there appear superfluid phases
with $p_x$ and $p_x+ ip_y$ symmetries. For $p_x+ ip_y$, there is a
transition between the weak-pairing phase (chemical potential $\mu>0$)
and the strong-pairing phase ($\mu<0$) as the resonance is swept from
positive to negative detuning\cite{gurarie}.
We propose to probe the properties of these phases by measuring the absorption spectrum, under
excitation of an atom in state $\down$  to a new internal state $\up$ via the
perturbation
\begin{eqnarray}
    H_{\mathrm{pert}}=\eta \int d^2 {\bm r}\, e^{i {\bm q}\cdot{\bm r}-i\omega t}\psi_\uparrow^\dag({\bm r})\psi_\downarrow({\bm r})+\mbox{h.c.},
\label{eq:perturb}
\end{eqnarray}
where $\eta$ sets the Rabi frequency of the transition.  We consider
$\down\to\up$ to be a hyperfine transition, driven by RF
absorption\cite{regal,chin} or a two-photon transition\cite{tewari};
measurement of the population of excited (or remaining) atoms is
achieved by separate absorption on an electronic transition.  We
denote the energy splitting between the states $\down$ and $\up $ by
$E_g$; for a hyperfine transition, $E_g\sim 10^9 \mbox{Hz}$, and we
set the net photon momentum transfer to zero, $q\to 0$.
The $\uparrow$-atoms will not participate in the resonant $p$-wave
interaction with the $\downarrow$-atoms, so for the most part we shall
consider the $\uparrow$-atoms to be free. However, we shall
discuss the effects of $s$-wave interspecies interactions.
We shall focus on a uniform gas in a quasi-two-dimensional
geometry, that is with confinement applied in the $z$-direction
such that the associated confinement energy is large compared to the
Fermi energy. This simplifies the analysis, allowing analytical
calculations of the absorption features. The three dimensional case is left for a future work.

When the interaction between atoms may be neglected (far from the Feshbach
resonance) and they form a simple Fermi gas, a transition of an atom
from $\down$ to $\up$ requires an energy of $E_g$, leading to a delta-function peak of the absorption at $\omega=E_g$. (We choose units for which $\hbar=1$.) In contrast, we find that in the
superfluid phases (close to the resonance) the absorption differs
strongly for weak- and strong-paired superfluids.  In the weak-pairing
phase the energy $E_g$ becomes a threshold energy for absorption,
with non-zero weight at $\omega = E_g$ and a continuous spectrum
above this (see Eq. (\ref{eq:abs-uniform})). In the strong pairing phase the threshold is shifted to
$E_g+2|\mu|$, with weight that is zero at $E_g+2|\mu|$ and
grows linearly with energy. The discontinuity in the absorption
spectrum in the weak-pairing phase unambiguously identifies it from strong pairing.

\begin{figure}[h]
  \begin{minipage}[t]{0.5\textwidth}
    \begin{center}
      \epsfig{file=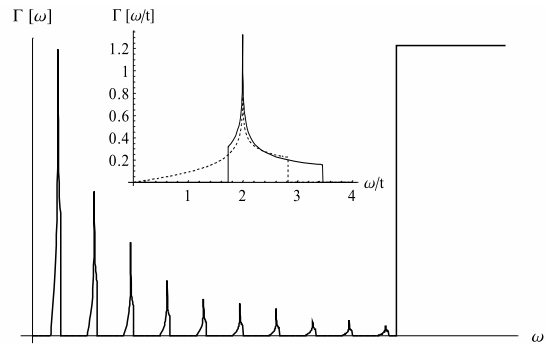, scale=0.38}
    \end{center}
  \end{minipage}
  \caption{Peak distribution associated with the Majorana
  modes on the vortices of the weak-pairing $p_x+ip_y$ phase.
  The first peak will be at
  $E\simeq E_g-\mu$, and other peaks will follow at spacings
  $\omega_c$ apart. Continuous absorption will be measured
  starting at $E\simeq E_g$. For the opposite sense of rotation, the weight of the peaks starts linearly and passes through a maximum. The inset shows the shape of a single peak for a square (dashed line) and triangular (solid line) lattice, showing a van-Hove singularity at
  $E=2t$ ($\Gamma$ is measured in units of $2\pi
  |\eta|^2 N_V\Theta_0/t$).
}
  \label{fig:absorption}
\end{figure}

In the weak-pairing $p_x \pm ip_y$ phase, the introduction of
vortices by rotation leads to the creation of a set of gapless
modes on the vortices.  In the presence of a vortex lattice
we find that \textit{a series of
equally spaced sub-gap peaks will be present in the absorption
spectrum}, see Fig. [\ref{fig:absorption}].  These
peaks are a result of transitions from the zero energy core states
of $\downarrow$-atoms to various states of $\uparrow$-atoms, and
have a weight that is linearly proportional to the number of
vortices $N_V$. The peaks start at $\sim E_g-\mu$, and their
weight decays monotonically in energy, becoming small well before
the onset of the continuum at $ E_g-\mu+\Delta$ ($\Delta$ is the
superfluid gap).  For a dense vortex lattice, in which there is tunneling
between the Majorana modes,
the peaks broaden to reveal a lineshape showing a van-Hove singularity,
see inset of Fig. [\ref{fig:absorption}].
For the same weak-pairing phase, with vortices of the
{\it opposite} sense of rotation, we find that the absorption in
the Majorana modes again gives rise to a set of narrow absorption
peaks. The peaks still start at $\sim E_g-\mu$, but now their
weight rises linearly in energy, passing through a maximum well before the
continuum. As we
shall discuss below, the difference in spectra for the two senses
of rotation is a signature of the fact that $p_x+ip_y$ breaks
time-reversal symmetry.

We now provide the details of the calculations. The rate of excitations between the two hyperfine states is
governed by the Golden Rule
\begin{equation}
    \label{eq:rate}
    \Gamma[\omega]=2\pi|\eta|^2\sum_{a b}
    |M_{ab}|^2\delta(E_{\downarrow,a}+E_{\uparrow,b}-\omega),
\end{equation}
where $E_{\downarrow,a}$ ($E_{\uparrow,b}$) is the energy required
to produce an $\downarrow$-excitation ($\uparrow$-excitation), and
$M_{ab}$ is the dimensionless matrix element (measured in units of
$\eta$) for producing excitations with quantum numbers $a$ and $b$
respectively. In the following we shall calculate
$\Gamma[\omega]$, first for a system at rest, then for a rotating
system.

We describe the superfluid phases
within the Bogoliubov-de-Gennes (BdG) approach. For a uniform system, the Hamiltonian is given by
\begin{eqnarray}
    \label{eq:general-hamiltonian}
    H=H_{0,\downarrow}+H_{\mathrm{pair},\downarrow}+H_{0,\uparrow}+E_g\left(\frac{N_\uparrow-N_\downarrow}{2}\right),
\end{eqnarray}
where $H_{0,s}=\int d^2 {\bm k}\,\psi_s^\dag({\bm
k})\left(\frac{{\bm k}^2}{2m}-\mu\right) \psi_s({\bm k})$,
$s=\uparrow,\downarrow$, $H_{\mathrm{pair},\downarrow}=\int d^2
{\bm k}\,\psi_\downarrow^\dag({\bm k})\v(k_x+i
k_y)\psi_\downarrow^\dag(-{\bm k})+\mathrm{h.c.}$. The values of $\mu$ and $\v$ may be
obtained from the mean-field calculations of
Refs.\onlinecite{cheng,gurarie}. The Hamiltonian for
$\downarrow$-atoms is diagonal in the Fock space of
$\alpha_{\bm{k}}=u_{\bm {k}} \psi_{\downarrow,\bm{k}}  + v_{\bm
{k}} \psi^\dag_{\downarrow,-\bm{k}}$ and $\alpha^\dag_{\bm{k}}
=u^*_{\bm {k}} \psi^\dag_{\downarrow,\bm{k}} +  v^*_{\bm {k}}
\psi_{\downarrow,-\bm{k}}$ where $u_{\bm k},v_{\bm k}$ are
solutions of the BdG equation,
$|u_{\bm{k}}|^2=\frac{1}{2}\left[1+\left(\frac{ {\bm
k}^2}{2m}-\mu\right)/E_{\downarrow,\bm k}\right]$ and
$|v_{\bm{k}}|^2=\frac{1}{2}\left[1-\left(\frac{ {\bm
k}^2}{2m}-\mu\right)/E_{\downarrow,\bm k}\right]$, and
$E_{\downarrow,\bm k}=\sqrt{\left(\frac{
{\bm k}^2}{2m}-\mu\right)^2+\v^2 {\bm k}^2}$ \cite{readgreen,gurarie}.
Treating the excited $\uparrow$-atoms as free,
$E_{\uparrow,{\bm k}}={\bm k}^2/2m-\mu$, a summation over ${\bm k}$
leads to the absorption spectrum \footnote{The unphysical
logarithmic divergence of the integrated intensity in
(\ref{eq:abs-uniform}) is a reflection of the need to introduce a
UV cut-off\cite{cheng} to regularise the theory for $\Delta_{\bm
k} \propto k_x+ik_y$.}
\begin{eqnarray}
    \label{eq:abs-uniform}
    \Gamma[\omega]=|\eta|^2\frac{S}{\v^2}\frac{\delta\omega+2\mu}{\left[1+\frac{\delta\omega}{m \v^2}\right]^2}\times\left\{\begin{array}{ll}
    \theta(\delta\omega), \quad & \mu>0 \vspace{2mm}\\
    \theta(\delta\omega-2|\mu|), \quad & \mu<0,
    \end{array}\right.
\end{eqnarray}
where $S$ is the area of the sample and $\delta\omega\equiv\omega-E_g$.
Non-zero interactions of the excited $\uparrow$-atom with the
$\downarrow$-atoms lead to a (mean-field) shift of both of these
spectra \footnote{There may also be a
renormalisation of the effective mass of the excited particle, and a
damping for $k\neq 0$ modes.}.

We now turn to discuss the system under rotation, in a regime
where a large vortex lattice is formed, $N_V \gg 1$. (The rotation
frequency $\Omega$ is then close to the trapping frequency
$\omega_T$.) We first examine a superfluid in the weak pairing $p_x+ip_y$
phase, and concentrate on the regime $\Omega\ll \mu\ll m \v^2$.

Projected to the low-energy sector of the Majorana modes,
the Hamiltonian for the $\downarrow$-atoms becomes a tight-binding
Hamiltonian of the form\cite{grosfeld}
\begin{eqnarray}
    \label{eq:hamiltonian-maj-tb}
    H_{\downarrow}=H_{0,\downarrow}+H_{\mathrm{pair},\downarrow}=i t
    \sum_{\langle ij\rangle}s_{ij}\gamma_{i\downarrow}\gamma_{j\downarrow},
\end{eqnarray}
where $s_{ij}=\pm$, in a way that corresponds to half a flux
quantum per plaquette for a square lattice, and a quarter of a
flux quantum per plaquette for a triangular lattice. Here $\gamma_i$
represents a Majorana fermion localized at ${\bm R}_i$. The
parameter $t$ describes the tunneling between the Majorana modes
of nearby vortices; this tunneling is assumed small, $t\ll \mu$,
which is valid if the separation between vortices, $\ell
\equiv\sqrt{1/(2m\Omega)}$, is large compared to the spatial
extent of the Majorana mode, $r_0$. The Majorana wave function
$g({\bm r})$ has an oscillating contribution within the vortex
core, and a decaying part outside the core. When $\mu\ll m \v^2$,
the size of the core can be estimated as $r_{\mathrm{core}}\sim
\frac{v_F}{m \v^2}\log^{-1/2}\frac{v_F}{\v}$, with $v_F$ being the Fermi velocity of the underlying fermionic gas,
while the decay length outside the core is $r_0=\v/\mu$. In the
limit of small $\mu$ we have $r_0\gg r_{\mathrm{core}}$, such that
we may approximate $g({\bm r})\approx\frac{e^{-r/r_0}}{\sqrt{2\pi
r_0 r}}$, with $r$ being the distance from the center of the vortex \cite{kopnin}.
The analysis of the Hamiltonian (\ref{eq:hamiltonian-maj-tb}) is
presented in \cite{kitaev2, grosfeld}.
For a square and triangular lattice the spectra are
$E_{{\scriptscriptstyle\square},{\bm k}}=2t\sqrt{\sin^2(a
k_x)+\sin^2(a
    k_y)}$ and $E_{{\scriptscriptstyle\triangle},{\bm k}}=\sqrt{2}t\sqrt{3+\cos(2 a k_x)-2\cos(a k_x)\cos(\sqrt{3} a
    k_y)}$ respectively\cite{grosfeld}.

The spectrum of the $\uparrow$-atoms in the rotating frame depends
on the strength of the $s$-wave interspecies interactions.  Since
this depends sensitively on the atomic species used, we shall
consider two limiting cases: For \textit{vanishing
interspecies interactions}, the $\uparrow$-atoms arrange into
Landau levels due to rotation, with the cyclotron frequency $\omega_c=2
\Omega \sim 2\pi \times 10^2\,\mathrm{Hz}$. (The confinement
frequency for $\uparrow$- and $\downarrow$-atoms is assumed the
same.) For \textit{strong interspecies
repulsion}, the low energy states of the $\uparrow$-atoms are
localized at the cores of vortices where the density of
$\downarrow$-atoms is low, and can be described by a tight-binding
Hamiltonian. The rotation then translates to a magnetic flux
threading the lattice plaquettes (the Azbel-Hofstadter problem
\cite{azbelhofstadter}).

Starting with the case of zero interspecies interactions
($V_\uparrow=0$), the Hamiltonian $H_{0,\uparrow}$ can be
diagonalized by expanding $\psi_\uparrow({\bm r})=\sum_{n p}\phi_{n
p}({\bm r})c_{np}$, where $\phi_{np}({\bm r})$ are wavefunctions for
the $n$'th Landau level, and $p$ is some quantum number related the
degeneracy of the Landau level. The rate of energy absorption is determined
by
\begin{widetext}
\begin{eqnarray}
    \label{eq:matrix-element}
    |M_{\alpha {\bm k},np}|^2=\frac{v}{N_V}\sum_{ij}\lambda^{\alpha*}_{z(j),{\bm k}}\lambda^{\alpha}_{z(i),{\bm k}}e^{i {\bm k}\cdot ({\bm R}_i-{\bm R}_j)}
    \int d^2 {\bm r}\,d^2 {\bm r}'\, e^{\frac{i}{2}\left[\sum_{m\neq i}\arg({\bm r}-{\bm R}_m)-\sum_{m\neq
    j}\arg({\bm r}'-{\bm R}_m)\right]}g_i({\bm r})g_j({\bm r}')\phi^*_{n p}({\bm r}')\phi_{n p}({\bm r}),
\end{eqnarray}
\end{widetext}
where $g_i({\bm r})=g({\bm r}-{\bm R}_i)$. The functions $\lambda^\alpha_{z(i),{\bm k}}$ are
listed in Ref.\onlinecite{grosfeld}, $z(i)=1,\ldots,v$ numbers the
vortices contained in a unit cell, and
$\alpha=\pm 1,\ldots,\pm v/2$ is a band index. Since $p$ enumerates degenerate states, we can now sum over it
using the addition theorem for Landau levels \cite{laikhtman}. We
keep only the leading term ${\bm R}_i={\bm R}_j$. Finally we
expand the phases in Eq. (\ref{eq:matrix-element}) to first order
in ${\bm r}-{\bm r}'$, and obtain $    \sum_p|M_{\alpha
    {\bm k},np}|^2=\frac{r_0^2}{\ell^2}\left[1+2n\left(\frac{r_0}{\ell}\right)^2\right]^{-2}$.
Thus the integrated absorption into the $n$'th Landau level falls
monotonically with increasing $n$.  Plugging the matrix elements into
Eq. (\ref{eq:rate}), we arrive at the result
\begin{eqnarray}
    \label{eq:rate-landau-levels}
    \Gamma[\omega]=2\pi\frac{|\eta|^2}{t} N_V \sum_n
    \Theta_n F\left[\frac{|\omega|-\omega_{n}}{t}\right],
    \label{absornoper}
\end{eqnarray}
where
$\Theta_n=\left(\frac{r_0}{\ell}\right)^2\left[1+2n\left(\frac{r_0}{\ell}\right)^2\right]^{-2}$,
$\omega_{n}=E_g-\mu+\omega_c(n+1/2)$, and $F$ is a
dimensionless function proportional to the density of states of
the band formed by the Majorana operators, which is nonzero over a range of order unity. It is explicitly
calculated below.

Using the results above, assuming $t\ll \omega_c$, we can find the
integrated absorption over a single Landau level
\begin{eqnarray}
    \label{eq:sum-rule1}
    \int^{\frac{\omega_c}{2}}_{-\frac{\omega_c}{2}}
    d(\omega-\omega_n)\Gamma=\pi |\eta|^2 N_V \Theta_n.
\end{eqnarray}
The integrated absorption over all frequencies for excitations from
the zero modes can be found using the completeness of the Landau
level functions $\sum_{np}|M_{\alpha {\bm k},np}|^2=\frac{1}{2}$, hence
$\int d\omega\Gamma[\omega]=\pi|\eta|^2 N_V/2$.

When interspecies interaction is strong, the
$\uparrow$-atoms feel a periodic potential $V_{\uparrow}$ due to
the vortex lattice structure. Due to depletion of
$\downarrow$-atoms at a vortex core, the potential $V_{\uparrow}$
possesses there a minimum, and may hold several bound states
$\phi_n({\bm r})$ for $\uparrow$-atoms. Several tight-binding
bands will be formed, labelled by $n$. The $\uparrow$-atoms feel
the same flux per plaquette as the Majorana fermions do.
Therefore, the Hamiltonian for a single band of tight-binding
atoms in the rotating frame may be written as
\begin{eqnarray}
    H_{0,\uparrow}=E'_n\sum_i c^\dag_{n,i\uparrow}c_{n,i\uparrow}+i
    t'_n\sum_{\langle ij\rangle}s_{ij}c^\dag_{n,i\uparrow}c_{n,j\uparrow},
\end{eqnarray}
where $t'_n$ is the tunneling matrix element for the
$\uparrow$-atoms, and $E'_n$ is an on-site energy.
Diagonalizing the Hamiltonian, one finds that the spectrum of the
$\uparrow$-atoms has the same ${\bm k}$-dependence
as that of the Majorana fermions.
The resulting absorption is given by Eq.
(\ref{eq:rate-landau-levels}) with $\Theta_n=|\int d^2 {\bm r}
g({\bm r})\phi_n({\bm r})|^2$, $\omega_{n}=E_g-\mu+E'_n$, $t\to
t+t'_n$.

We now proceed to calculate the function $F$ appearing in Eq.
(\ref{absornoper}). This function is $F[E/t]\equiv t\rho[E]/N_V$,
with $\rho(E)$ being the density of states of the Majorana
fermions. Starting with a square lattice, $F[x]$ is
\begin{eqnarray}
    F_{\scriptscriptstyle{\square}}[x]=\frac{8 x}{(2\pi)^2}\int_C
    \frac{d y}{\sqrt{x^2-y^2}\sqrt{4-x^2+y^2}\sqrt{4-y^2}}.
\end{eqnarray}
For $0<x<2$, $C=\left[0,x\right]$, while for $2<x<2\sqrt{2}$
$C=\left[\sqrt{x^2-4},2\right]$. We note the diverging density of
states at $E=2t$, and the discontinuity at the top of the band
$E=2\sqrt{2}t$. For a triangular lattice, we
again find a logarithmic divergence near $E=2t$ and
discontinuities in the response at the top and the bottom of the
band, $E=2\sqrt{3}t$ and $E=\sqrt{3}t$ respectively.

For the {\it opposite} sense of rotation, the vortex lattice
consists of an array of vortices of opposite circulation. These
vortices also have zero modes\cite{grvortex}.  However, the
Majorana modes now have a non-trivial dependence around the vortex
center, $g_i({\bm r})\propto e^{-i\arg{({\bm r}-{\bm R_i})}}$.
(We again assumed $r_0\gg r_{\rm core}$.)  As a consequence, the
integrated weight of the absorption from the Majorana mode into
the $n$'th Landau level is proportional, for small $n$, to
\begin{equation}
    \sum_p |M|^2 =
    \frac{9\pi}{64}\left(\frac{r_0}{\ell}\right)^4
    \left(n+\frac{1}{2}\right).
\end{equation}
The weight passes
through a maximum at a frequency of about $\frac{\mu^2}{m \v^2}$
above the onset at $E_g -\mu$.

Thus we find that in the weak pairing $p_x+ ip_y$ phase, the
spectrum depends strongly on the sense of rotation. This
dependence may be probed by preparing the system in the $p_x+ip_y$
phase and comparing spectra for rotations of the two senses.
Alternatively, since what matters is the {\it relative} sense of
rotation as compared to the internal angular momentum of the order
parameter, the dependence could be tested by comparing the spectra
of the $p_x+ip_y$ and $p_x-ip_y$ phases for the same sense of
rotation. In a stationary fluid, the two phases are equally likely
to be formed
\footnote{There is also the possibility of formation of an inhomogeneous
state with domains of the two phases.}.  If for a given sense of
rotation absorption spectra of two types are randomly observed for
different realizations of nominally the same parameters, it would
be a direct demonstration of time-reversal symmetry breaking.
We note that the results may depend on how the state was formed (apply
rotation and then cool, or cool and then apply rotation)\footnote{In a
rotating system time-reversal symmetry is explicitly broken, so, in a
``rotation cooled'' experiment, one phase may form in preference over
the other.}.

The observation of absorption that originates from Majorana modes
would also distinguish the time-reversal symmetric $p_x$ phase
from the phase of $p_x + i \beta p_y$, with a real $\beta\ne 0$,
where time-reversal symmetry is broken. The latter may appear for
an anisotropic Feshbach resonance\cite{cheng,gurarie}. The
distinction arises, as we now show, from the presence of Majorana
modes for all $\beta\ne 0$, and their absence for $\beta =0$.
Solutions to the BdG equations come in pairs of $\pm E$. When a
vortex carries a localized single zero energy state, as it does
for $\beta=1$, the only ways the energy of that state may shift
due to a change of $\beta$ are through mixing with other zero
energy states (on other vortices, or on the sample's edge) or when
the bulk gap closes. When vortices are far away from one
another and from the edge, the zero energy state exists as long as
the gap does not close, i.e., as long as $\beta\ne 0$
\cite{ivanov}\footnote{We note that the anisotropic order parameter
of the $p_x+i\beta p_y$ phase will choose a preferred direction for tunneling between vortices.}.

To conclude, our results show that the absorption spectrum can be
used to distinguish the various $p$-wave superfluid phases of an
atomic Fermi gas, and to probe the breaking of time reversal
symmetry. Furthermore, for a rotating superfluid in the weak
pairing $p_x+ip_y$ phase, the absorption shows a series of sub-gap
peaks, of unique shape and distribution, which are the direct
consequence of the zero-energy Majorana modes on the vortices.
The most favourable parameters for the observation of these
sub-gap peaks are $\mu\ll m \v^2$  (to avoid finite energy
bound states in the vortex core), and temperature  $T\ll \mu$ (to
avoid thermally excited quasi-particles which may wash out the
absorption peaks). A practical way to observe the peaks would be
by comparing measurements of the absorption for increasing
rotation frequency. This will vary both the strength of the peaks
(due to the increase of $N_V$) and their spacing (due to the
increase of $\omega_c$).

We acknowledge financial support from the US-Israel BSF (2002-238) and
the ISF, and from EPSRC Grant No. GR/S61263/01 and the ICAM Senior
Fellowship Programme (NRC).

% ----------------------------------------------------------------

\end{document}